\documentclass[aps,amssymb,pra,twocolumn,showpacs,floatfix]{revtex4}

\usepackage{epsf}
\usepackage{amsmath}
\usepackage{graphicx}
\begin{document}
         
\title {Subatomic mechanism of the oscillatory magnetoresistance in superconductors}

\author{Boris I. Ivlev}

\affiliation
{Instituto de F\'{\i}sica, Universidad Aut\'onoma de San Luis Potos\'{\i},
San Luis Potos\'{\i}, 78000 Mexico\\}

\begin{abstract}
In the recent experiments \cite{MIL} the unusual oscillatory magnetoresistance in superconductors was discovered with a periodicity essentially independent on magnetic field direction and even 
material parameters. The nearly universal period points to a subatomic mechanism of the phenomenon. This mechanism is related to formation inside samples of subatomically thin ($10^{-11}cm$) 
threads in the form of rings of the interatomic radius. Electron states of rings go over into conduction electrons which carry the same spin imbalance in energy as rings. The imbalance occurs due 
to spin interaction with the orbital momentum of the ring. The conductivity near $T_c$ is determined by fluctuating Cooper pairs consisting of electrons with shifted energies. Due to different 
angular momenta of rings these energies periodically depend on magnetic field resulting in the observed oscillatory magnetoresistance. Calculated universal positions of peaks 
$(n+1/2)\Delta H$ ($\Delta H\simeq 0.18T$ and $n=0,1,2...$) on the $R(H)$ curve are in a good agreement with measurements.

\end{abstract} \vskip 1.0cm


\maketitle

\section{INTRODUCTION}
\label{review}
In normal metals there are well known Shubnikov-de Haas oscillations of resistance in a high magnetic field \cite{ABR}. In superconductors 
oscillations of magnetoresistance also occur and they do not require such high magnetic fields. The scale of oscillations in many cases is related 
to the magnetic flux quantum $\Phi_0=\pi c\hbar/e$. Oscillations in magnetic field in superconductors has a long history \cite{BARD,LITTL1,LITTL2,TINK,LUB,KOP1,KOP2,RAMM,CHAIK,GOLD,BISH,BLO,MOSH1,
GRIES,MOSH2,HER,MOSH3,ABI,PEE,KOG,UST,MOSH4,PAT1,PAT2,BERD1,BERD2,YESHUR,BERD3}. Oscillations can occur in Josephson junctions, as Little and Parks effect, due to variation of number of phase slips
centers and lines, in layered and granular superconductors, due to artificial geometrical restrictions, etc.

Every oscillation effect in superconductors has clear and well studied background. At first sight, unlikely another oscillation phenomenon may exist whose mechanism is mysterious, that is 
outside the circle of known effects. Nevertheless the unusual oscillating magnetoresistance, experimentally observed in \cite{MIL}, stays well apart. Its underlying mechanism cannot be reduced to 
a combination of known effects since the periodic positions of $R(H)$ peaks are universal, that is material independent. In layered compounds observed peak positions are independent on direction of 
the magnetic field.

The observed properties are compatible with a subatomic mechanism which controls conduction electrons in a relatively large volume. This would provide material independence since subatomic states 
have no resemblance to atoms of the solid. Such a construction looks paradoxical for two reasons: (i) formation mechanism of electron states with subatomic size is unclear and (ii) it is unclear 
how a subatomically small state can control conduction electrons responsible for macroscopic properties. 

The mechanism of formation of subatomic states in condensed matter is unusual. Under the electron-photon interaction the electron ``vibrates'' with the mean displacement 
$\langle\vec u\rangle=0$ and the mean squared displacement $r^{2}_{T}=\langle u^2\rangle$ where $r_T\sim 10^{-11}cm$ \cite{WEL,MIGDAL,KOL}. This is the fluctuation spreading in addition to the 
usual quantum mechanical uncertainty \cite{LANDAU5}. For example for the harmonic oscillator $m\Omega^2R^2/2$ the total mean squared displacement is $3\hbar/2m\Omega+\langle u^2\rangle$. In this 
language the ``vibrating'' electron probes various parts of the potential and therefore changes its energy (Lamb shift) \cite{LANDAU3}. 

In quantum mechanics, regardless of a form of the potential, the electron wave function can be singular along the $z$ axis as $\psi\sim\ln r$ where $r^2=x^2+y^2$. In this case the kinetic energy 
term $-\hbar^2\nabla^2/2m$ is singular as $\delta(\vec r)$. To compensate this singularity in the wave equation the artificial $\delta(\vec r)$ should be added as a formal potential well. Such 
singularity source is absent in reality and therefore the singular state does not exist even formally. 

However the singularity source appears on short distances, $10^{-16}cm$, from the singularity line due to the mechanism of electron mass formation \cite{ENG,HIG,GUR}. Then under the interaction 
with photons, electron ``vibrations'' smear the singularity into the thread of the radius  $r_T$ along the $z$ axis. Within the thread the term $-\hbar^2\nabla^2/2m$ goes over into 
$\hbar^2/mr^{2}_{T}$. As shown in \cite{IVLEV4,IVLEV2}, that large kinetic energy is compensated by the counter-term that can be interpreted as anomalous well along the thread. This term is formed 
by the variation in space of zero point electromagnetic energy on the distance $10^{-11}cm$ around the thread. 

The resulting state is smooth in space and therefore physical. It is localized within the thread of the thickness $r_T$. The thread is not necessary linear. The subatomically thin thread can be in 
the form of a ring of the atomic radius. In a metal the role of such rings is unusual due to orbital momenta $l_z$ of the ring along the $z$ axis perpendicular to the ring plane. Close to the thread 
(on the Compton length) relativistic effects are strong resulting in $j_zj_z$ coupling analogous to $jj$ coupling in some atoms. Due to the interaction with photons, the state of the Fermi energy is 
split by two ones, with $l_z=j_z+1/2$ and  $l_z=j_z-1/2$. This is similar to the Lamb split in hydrogen atom where, instead of $z$ component, total momenta are involved due to spherical symmetry.  

Those states, with energy split for opposite spins, continue from the thread ring to larger distances going over into conduction electrons of $E_F$ energy. The narrow stiff region near the thread 
plays a role of a boundary condition for conduction electrons moving in crystal field. This keeps electrons with opposite spins in the volume (on the length of spin-orbit relaxation) to be 
separated by discrete Lamb energies. The driving force for that spin imbalance state is thread rings distributed in the volume.

The spin imbalance state of conduction electrons, with discrete energy splits for opposite spins, influences the Cooper pairing condition in the fluctuation region close to $T_c$. The 
fluctuation correction to the resistance of normal metal is determined by the fluctuation propagators which depend on energy shifts of different spin states. These energy shifts can be subsequently 
turned to zero by the external magnetic field. Therefore the resulting $R(H)$ dependence becomes oscillating as in experiments. 

Actually the measurements \cite{MIL} probe the spin imbalance state in the volume of a metal. In contrast to known mechanisms of $R(H)$ oscillations, the spin imbalance mechanism does not depend on
macroscopic inhomogeneities of samples. 

In Sec.~\ref{mechanism} usual oscillation are analyzed. In Secs.~\ref{anom}, \ref{shape}, and \ref{states} the mechanism of anomalous states is studied. In Sec.~\ref{spin} spin imbalance states are 
introduced. In Sec.~\ref{cooper} effects on the fluctuation region are investigated.
\section{UNUSUAL OSCILLATIONS}
\label{mechanism}
In the film with the artificial periodic two-dimensional structure the oscillatory magnetoresistance is due to effects generic with Little-Parks 
phenomenon. See for example \cite{YESHUR}. The period of $H$ oscillations was determined as
\begin{equation}
\label{10}
\Delta H=\frac{\Phi_0}{\sigma}\,,
\end{equation}
where $\sigma$ is the unitary cell area of the structure. 

An oscillatory behavior of resistance is also possible when the superconducting sample consists of natural grains of the typical size 
$\sqrt{\sigma}$ \cite{HER,PAT2}. In this case the typical distance between peaks of $R(H)$ is also determined by the geometrical condition
(\ref{10}). When the magnetic field is perpendicular to the film surface $width\times length$ the area $\sigma$ is determined by grain structure 
on that surface. When the field is perpendicular to the side surface $width\times thickness$ the corresponding $\sigma$ is less (a larger 
$\Delta H$) due to the geometrical restriction by the finite thickness. This clear property, dependence of $\Delta H$ on $\vec H$ direction, was 
observed in experiments \cite{PAT2}.

It is also clear that in a naturally disordered sample grain sizes cannot be equal resulting in a perfect periodicity of $R(H)$ with the period 
(\ref{10}). Analogously the perfect periodicity is not expected with the same $\Delta H$ in samples of different materials and differently 
manufactured. The two main features distinguish the oscillatory magnetoresistance observed in \cite{MIL}: independence of $\Delta H$ on 
$\vec H$ direction and independence of $\Delta H$ on sample choice. 
\subsection{Independence of $\Delta H$ on $\vec H$ direction}
In experiments \cite{MIL} the distance $\Delta H$ between maxima of $R(H)$ is constant with the accuracy of $5\%$. In addition, $\Delta H$ is the same for all $\vec H$ directions. 
This points to a different mechanism of the oscillations of magnetoresistance compared to the 
geometrical origin (\ref{10}). Indeed, films thickness in \cite{MIL} is smaller than $\sqrt{\sigma}$ and therefore projections of grain areas to 
the side surface and upper one cannot be equal. Therefore the nature of magnetoresistance oscillations in that case is not of geometrical origin 
(\ref{10}), as in Refs.~\cite{YESHUR,HER,PAT2}, but qualitatively different. 
\subsection{Independence of $\Delta H$ on sample choice}
$\Delta H$ was revealed to be equal for ${\rm Sr}_{1-x}{\rm La}_{x}{\rm CuO}_{2}$ and ${\rm Y}_1{\rm Ba}_2{\rm Cu}_3{\rm O}_7$ \cite{MIL}. This 
independence of $\Delta H$ also looks surprising and says against the geometrical origin (\ref{10}). Otherwise samples of different materials
have to have identical and perfectly periodic grain structure.
\subsection{Various mechanisms}
The stable feature (independence of macroscopic properties) of magnetoresistance oscillations says about a subatomic mechanism as mostly probable 
one. In geometrical effects, generic with Little-Parks one, magnetoresistance oscillations occur due to periodic in $H$ diamagnetic pair breaking. 
Associated $\Delta H$ is analogous to (\ref{10}) and is not universal with respect to different $\vec H$ directions. In contrast, the mechanism of 
paramagnetic pair breaking is promising since it is not geometrical and therefore is expected to provide no angular $\vec H$ dependence. But 
such a mechanism should also result in the oscillatory $R(H)$. It is not clear a priori how it can be.

In this paper we investigate that phenomenon. It is shown that anomalous electron-photon states are likely responsible for the observations \cite{MIL}. 
\section{ANOMALOUS ELECTRON STATES}
\label{anom}
Since the observed oscillations are universal, this drives to analyze a possible subatomic mechanism to avoid a dependence on material choice and $\vec H$ direction. The subatomic mechanism, 
considered from usual atomic distances, look like a singularity of the wave function. At shorter (subatomic) distances the singularity has to be washed out within the certain small scale. In this 
section we show that such anomalous states can really exist.
\subsection{What follows from wave equation}
\label{anomA}
First, we consider an electron without an interaction with any (in particular, electromagnetic) fluctuating field to be included as the second step. Such electron is described by a quantum 
mechanical wave equation in some static potential $U(\vec R)$. When the potential is physically smooth, the Schr\"{o}dinger equation in the space $\vec R=\{\vec r,z\}$ 
\begin{equation}
\label{M0}
\left(-\frac{\hbar^2}{2m}\nabla^2+U\right)\psi=E\psi
\end{equation}
can have the singular solution, which is $\psi\sim\ln r$ at $r\rightarrow 0$, extended along the $z$ axis. This solution requires the singularity source $\delta(\vec r)$ in the right-hand side of 
(\ref{M0}). This source is absent in this equation and therefore that singular solution does not exist even formally. 

However it is not clear whether such source appear under the reduction of $r$ when the formalism (\ref{M0}) is not valid. The Schr\"{o}dinger description (\ref{M0}) of that singularity holds at 
$r_c<r$ where $r_c=\hbar/mc\simeq 3.86\times 10^{-11}cm$ is the electron Compton length. At $r<r_c$ one has to use the Dirac quantum mechanics for the bispinor $\psi=(\varphi,\chi)$ where $\varphi$ 
and $\chi$ are two spinors satisfying the equations for free electron \cite{LANDAU3} 
\begin{equation}
\label{M1}
(\varepsilon+i\hbar c\vec\sigma\nabla)\varphi=mc^2\chi,\hspace{0.4cm}(\varepsilon-i\hbar c\vec\sigma\nabla)\chi=mc^2\varphi.
\end{equation}
Here $\varepsilon$ is the total relativistic energy and $\vec\sigma$ are Pauli matrices. In equations (\ref{M1}) the gradient terms are large and the static potential is neglected since it is less 
than $mc^2$. Equations (\ref{M1}) follow from the Dirac Lagrangian \cite{LANDAU3}
\begin{equation}
\label{M2}
L=i\hbar c{\bar\psi}\gamma^{\mu}\partial_{\mu}\psi-mc^2{\bar\psi}\psi,
\end{equation}
where $\gamma^{\mu}$ are Dirac matrices, $\bar\psi=\psi^{*}\gamma^0$ is the Dirac conjugate, and the partial derivatives are $\partial_{\mu}=(\partial/\partial ct,\nabla)$. 

It follows from (\ref{M1}) that
\begin{equation}
\label{M3}
(\varphi-\chi)=-\frac{i\hbar c}{\varepsilon+mc^2}\vec\sigma\nabla(\varphi+\chi).
\end{equation}
To be specific, one can choose the spinor $(\varphi+\chi)$ in the form
\begin{equation}
\label{M4}
(\varphi+\chi)=\frac{1}{\sqrt{2}}{1\choose 1}F,
\end{equation}
where $F$ satisfies the equation
\begin{equation}
\label{M5}
\left(-\nabla^2+\frac{m^2c^2}{\hbar^2}\right)F=\frac{\varepsilon^2}{\hbar^2c^2}F.
\end{equation}
This equation is similar to (\ref{M0}) (with no potential) when $\varepsilon=mc^2+E$ and the energy $E$ is small compared to $mc^2$. 

As follows from Eq.~(\ref{M5}), at small $r$ the term with $\nabla^2$ dominates and the singular solution $F\sim\ln r$ also requires the singularity source $\delta(\vec r)$ which is absent in 
(\ref{M5}). Therefore our attempt to naturally get a singularity source at smaller $r<r_c$ failed. The singular solution, continued to the region $r<r_c$, does not exist as in the Scr\"{o}dinger 
formalism. Note that at $r<r_c$ the combination $(\varphi-\chi)\sim r_c/r$ dominates but at $r_c<r$ the term $(\varphi+\chi)\sim\ln r$ is the principal one related to the Scr\"{o}dinger 
equation. 
\subsection{Beyond wave equation}
\label{anomB}
Below we analyze what happens to the singularity on much shorter distances $r$ compared to the electron Compton length $r_c$. Also one should specify a physical origin of that shorter distance. 

According to the Standard Model, masses of electron, other leptons, $W^{\pm}$ and $Z$ weak bosons, and quarks are generated by the Higgs field \cite{ENG,HIG,GUR}. Electron acquires its mass through
the connection between the fermion field $\psi$, which includes electrons, and the Higgs field $\phi$. Instead of the electron mass in the Lagrangian (\ref{M2}) the term, connecting $\psi$ and 
$\phi$, appears. One should formally substitute $mc^2\rightarrow G\phi$ where $G\sim m/\mu\sim 10^{-5}$ and $\mu\sim 100GeV/c^2$ is the mass of the Higgs boson. So the last part in the Lagrangian 
(\ref{M2}) is $-G\bar\psi\phi\psi$ which is called the Yukawa term. 

The Higgs field $\phi=v+h$ contains the fluctuating part $h$ for which $\langle h\rangle=0$. Therefore the electron mass $m$ is determined by the expectation value $v$ of the Higgs field
\begin{equation}
\label{M6}
mc^2=Gv.
\end{equation}

Besides the generation of electron mass, the Yukawa $\phi$ depending term in the Lagrangian also influences the Higgs field. As above, we consider the problem without fluctuating gauge fields 
$W^{\pm}_{\mu},Z_{\mu},A_{\mu}$ ($A_{\mu}$ relates to photons) and the fluctuating part $h$ of the Higgs field. In this case the expectation value $v$ of the Higgs field obeys the equation
\cite{ENG,HIG,GUR}
\begin{equation}
\label{M7}
\hbar^2c^2\nabla^2v+\mu^2v-v^3=\frac{\hbar^3c^3}{2}G\langle\bar\psi\psi\rangle.
\end{equation}
The right-hand side of (\ref{M7}) can be calculated according to Dirac quantum mechanics \cite{LANDAU3}
\begin{equation}
\label{M8}
\bar\psi\psi =\varphi^{*}\chi+\chi^{*}\varphi=\frac{1}{2}\left(|\varphi+\chi|^2-|\varphi-\chi|^2\right)
\end{equation}
since fluctuating fields are absent. 

Eq.~(\ref{M7}), producing the finite expectation value of the Higgs field, reminds the Ginzburg-Landau equation. The peculiarity of (\ref{M7}) is its right-hand side mainly determined by the singular
part (\ref{M3}) in Eq.~(\ref{M8}). That part is essentially coordinate dependent that makes $v$ also a function of $r$. According to (\ref{M6}), the mass in the Lagrangian (\ref{M2}) becomes
variable in space. One can easily show that the relations (\ref{M3}) and (\ref{M4}) in this case remain the same but Eq.~(\ref{M5}) now reads 
\begin{equation}
\label{M9}
\left(-\nabla^2+\frac{\nabla mc^2}{\varepsilon+mc^2}\nabla+\frac{m^2c^2}{\hbar^2}\right)F=\frac{\varepsilon^2}{\hbar^2c^2}F.
\end{equation}
In Eq.~(\ref{M7}) the expectation value becomes variable $v\rightarrow v+\delta v(r)$. According to (\ref{M6}), the electron mass is also variable $m\rightarrow m+\delta m(r)$. As follows from
(\ref{M7}), (\ref{M8}), and (\ref{M3}), 
\begin{equation}
\label{M10}
\left(-\nabla^2+\frac{2}{R^{2}_{c}}\right)\delta m=G^2\frac{\hbar^3c}{4}\left(\frac{\nabla F}{\varepsilon+mc^2}\right)^2,
\end{equation}
where $R_c=\hbar/\mu c\sim 10^{-16}cm$ is the Compton length of the Higgs boson. In Eq.~(\ref{M9}) at $r<r_c$ only gradient terms are significant. It follows that the expression
$r(\partial F/\partial r)/(\varepsilon+mc^2)$ is a constant. In the limiting cases \cite{IVLEV4}
\begin{equation}
\label{M11}
\frac{\delta m(r)}{m}\sim G^2
\begin{cases}
R^{2}_{c}/r^2,&R_c<r,\\
-(\ln R_c/r)^2,&r<R_c.
\end{cases}
\end{equation}
The variable mass correction is localized at $r\lesssim R_c$ and always small. The main $r$-dependence $F\sim\ln R_c/r$ is added by the correction proportional to $G^2(\ln R_c/r)^3$ \cite{IVLEV4}. 
The electron density $n\sim (\partial F/\partial r)^2\sim 1/r^2$ at $r\ll r_c$. 

We consider the topological case when the phase of the wave function remains the same after completing the circle around the $z$ axis. Generally speaking, in this process the wave function can be
multiplied by $\exp(2i\pi\nu)$. In this case the electron density is $n\sim 1/r^{2+2\nu}$ at small $r$. This situation requires further studies. 

Without the $\nabla m$ term in (\ref{M9}) it would be $\nabla^2F\sim\delta(\vec r)$ with the non-existing singularity source in the right-hand side. After the subsequent average on fluctuating
fields (Sec.~\ref{anomC}) the $\delta$ function would be smeared within a finite region resulting in the non-existing term extended in space. In contrast, with the $\nabla m$ term in (\ref{M9}) the
kinetic part $\nabla^2F\sim G^2/r^2$ exists at any $r\rightarrow 0$ and after the average it goes over into the smooth part that is physical. In other words, the $\nabla m$ term provides the 
singularity source which is localized at short distances $r\lesssim R_c$. These distances correspond to the condition $1/r^{2}_{c}<G^2/r^2$ of domination of $\nabla m$ term in Eq.~(\ref{M9}). 
\subsection{Smearing of the singularity}
\label{anomC}
The solution obtained remains singular until fluctuations of gauge fields $W^{\pm}_{\mu},Z_{\mu},A_{\mu}$ and of the field $h$ enters the game. These fluctuations are expected to wash out the 
singularity within the certain radius $r_T$ around the $z$ axis. Masses of the fields $W^{\pm}_{\mu},Z_{\mu}$ and $h$ are large, about of $100GeV/c^2$. For this reason, fluctuations of these 
fields result in a less fluctuation length compared to fluctuations of the massless photon field $A_{\mu}$. Therefore for study of singularity smearing one can account for solely the electron-photon
interaction.

To generally understand how the singularity is washed out let us account for photons by implementation of the multi-dimensional quantum mechanics where photons are the infinite set of harmonic
oscillators \cite{LANDAU3}. See also \cite{LEG,IVLEV2}. The total eigenenergy of the stationary state is
\begin{equation}
\label{M12}
E_{tot}=E(\vec r,z)+\sum\frac{\hbar\omega}{2}-\left(\sum\frac{\hbar\omega}{2}\right)_0,
\end{equation}
where the first term relates to the electron part which also includes the interaction with photons. The last term is the zero point energy of photons in absence of the electron. A dependence on 
$\vec r$ and $z$ in the second term of (\ref{M12}) comes from a spatial dependence of the photon density of states.

Far away from the $z$ axis the state is hardly violated by the interaction with photons due to smallness of $e^2/\hbar c$. Because of locality of the system, described by differential equations, 
one can track the exact stationary solution (with the total energy $E_{tot}$) in the multi-dimensional space from large to small $r$. The state, continued from the infinity, comes to the singularity 
at the new position $\vec r=\vec u$ which depends on a choice of photon degrees of freedom. The electron density, calculated in Sec.~\ref{anomB}, now becomes
\begin{equation}
\label{M13}
n\sim\frac{1}{(\vec r-\vec u)^2}\,.
\end{equation}
Each fixed set of electromagnetic variables specifies in three-dimensional space the singularity curve ($\vec u(z)$ in (\ref{M13})) localized around the $z$ axis. Without the electron-photon 
interaction $\vec u=0$ as in Sec.~\ref{anomB}. An average on photon degrees of freedom leads to a superposition of states with various singularity curves. The resulting state is smooth. It recalls
the thread of the certain thickness $r_T$ along the $z$ axis. Below this thickness is determined.
\subsubsection{Lamb shift}
Suppose that in the three-dimensional potential well $U(R)$ ($R^2=r^2+z^2$) the ground state energy of the electron is $E$ in the absence of the interaction with photons. Under this interaction
the electron ``vibrates'' with displacements $\vec u$. The related mean displacement $\langle\vec u\rangle=0$ but the mean squared displacement $r^{2}_{T}=\langle u^2\rangle$ is finite. The
effective potential can be estimated as \cite{WEL,MIGDAL,KOL}
\begin{equation}
\label{101}
\langle U(|\vec R-\vec u|)\rangle\simeq U(R)+\frac{\langle u^2\rangle}{6}\nabla^2U(R).
\end{equation}
The quantum mechanical perturbation, due to the second term in (\ref{101}), leads to the eigenenergy deviated from $E$ by the Lamb shift $\delta E_L$ \cite{LANDAU3}
\begin{equation}
\label{102}
\delta E_L=\frac{\langle u^2\rangle}{6}\int\psi^*(\vec R)\nabla^2U(R)\psi(\vec R)d^3R.
\end{equation}
When the potential is the harmonic oscillator $U(R)=m\Omega^2R^2/2$, or it is close to it at small $R$, the total mean squared displacement is 
\begin{equation}
\label{102a}
\langle R^2\rangle=\frac{3\hbar}{2m\Omega}+\langle u^2\rangle,
\end{equation}
where the first part is the usual quantum mechanical uncertainty. One can calculate \cite{WEL,MIGDAL,KOL,IVLEV2}
\begin{equation}
\label{103}
r^{2}_{T}=\langle u^2\rangle=\frac{2r^{2}_{c}}{\pi}\frac{e^2}{\hbar c}\ln\frac{mc^2}{\hbar\Omega}\,.
\end{equation}
It follows that $r_T\sim 10^{-11}cm$. The Lamb shift (\ref{102}) of the ground state energy, with the result (\ref{103}), is valid with the logarithmic accuracy and it can be obtained  without the 
full machinery of quantum electrodynamics just applying non-relativistic quantum mechanics \cite{WEL,KOL,IVLEV2}. To go beyond the logarithmic accuracy the non-relativistic approach is not 
sufficient.

The results (\ref{102}) and (\ref{103}) are applicable to hydrogen atom where $U(R)=-e^2/R$, $\nabla^2U=4\pi^2e^2\delta(\vec R)$, and $|\psi(0)|^2=(me^2/\hbar^2)^3/\pi$. In this case one should 
substitute $\hbar\Omega$ in (\ref{103}) by Rydberg energy \cite{WEL,MIGDAL,KOL}. Eq.~(\ref{102}) produces the Lamb shift of the ground state of hydrogen atom
\begin{equation}
\label{104}
\delta E_L=\frac{8mc^2}{3\pi}\left(\frac{e^2}{\hbar c}\right)^5\ln\frac{\hbar c}{e^2}\,,
\end{equation}
that coincides with the exact (with the logarithmic accuracy) result following from quantum electrodynamics \cite{LANDAU3}. 

The Lamb shift of levels of the harmonic oscillator $U(R)=m\Omega^2R^2/2$ is
\begin{equation}
\label{105}
\delta E_L=\frac{m\Omega^2}{2}\langle u^2\rangle,
\end{equation}
where the mean squared displacement is given by Eq.~(\ref{103}). 
\subsubsection{Smearing of the singularity}
We see that electron ``vibrations'' due to its interaction with photons results in the typical fluctuation length $r_T$. The singularity along the $z$ axis is washed out within the thread 
(along the $z$ direction) of the subatomically small radius $r_T\sim 10^{-11}cm$. This thread state can be called {\it anomalous electron state}. 

One should emphasize that smearing of the singularity, within the finite radius $r_T$, occurs solely when the $z$ axis coincides with the equilibrium position of the electron. This corresponds to 
the potential $m\Omega^2r^2/2$ at small $r$. For a free electron the thread radius $r_T=\infty$ since $\Omega=0$. In this case anomalous state does not exist. Instead there is the usual Lehmann 
representation of the electron propagator in quantum electrodynamics \cite{LANDAU3}. In other words, anomalous state is impossible for free electron.

The direct average of the electron density (\ref{M13}) formally results in the logarithmic divergence at small arguments. The direct average of higher spatial derivatives of $n$ results in even
stronger divergences. For this reason, it is convenient to average the number of electrons which are at the interval between $r$ and $r_c$
\begin{equation}
\label{M14}
N(r)=2\pi a\int^{r_c}_{r}n(r_1)r_1dr_1\sim \ln\frac{r_c}{r},
\end{equation}
where $a$ is the length of the thread. 

After the average $\langle N(r)\rangle$ becomes a smooth function of $r$ with the typical scale $r_T$. Its derivative with respect to $r$ produces the physically smooth electron density with the 
same typical scale in $r$. This density has the peak at the thread position $n(r_T)\sim n(r_c)r^{2}_{c}/r^{2}_{T}\sim n(r_c)\hbar c/e^2$. 
\subsection{Origin of the $MeV$ well}
\label{anomD}
The peak of the electron density at $r\lesssim r_T$ can be interpreted as enhancement of the electron kinetic energy $\hbar c/r_T$ at that region. Formally this corresponds to the domination of 
the kinetic term $\nabla^2F$ in Eq.~(\ref{M9}). 

On the other hand, we continue the exact stationary state of the multi-dimensional system (Sec.~\ref{anomC}) with the energy (\ref{M12}) from large $r$. At fixed $E_{tot}$ various photon field 
configurations lead to the above local enhancement of the first term in (\ref{M12}). This enhancement has to be compensated by the local reduction of the second term in (\ref{M12}) just to keep the 
same $E_{tot}$. Therefore the spatial redistribution of the photon density of states in (\ref{M12}) is adjusted to produce the certain well, along the $z$ axis, localized at $r\lesssim r_T$ around 
this axis. The depth of this well, formed by the reduction of the vacuum energy, is 
\begin{equation}
\label{107}
U_0\sim\frac{\hbar c}{r_T}\sim mc^2\sqrt{\frac{\hbar c}{e^2}}\,.
\end{equation}
One estimates $U_0\sim 1MeV$. As follows from (\ref{107}), $U_0$ cannot be obtained from the perturbation theory on $e^2/\hbar c$ despite this parameter is small. The reason is that the electron 
density is proportional to $1/(\vec r-\vec u)^2$ where the both displacements are of the same order at $r\lesssim r_T$. 

So the states in the well relate to the non-perturbative approach and they are exact. This means that each state is non-decaying, ${\rm Im}E_{tot}=0$. Another property is that one can take any 
energy $E_{tot}$ and arrive to the thread state. Therefore the spectrum of states in the well is continuous and non-decaying. This contrasts with a usual potential well which is fixed and is not 
adjusted to each electron state. The continuous non-decaying spectrum in a well in presence of a continuous medium is not forbidden in nature. Such spectrum was revealed in Ref.~\cite{IVLEV3} on 
the basis of the exact solution. 

The similar well creation occurs, for example, in attraction of two hydrogen atoms at large distances \cite{CAS,LANDAU3}. This Casimir (van der Waals) attraction is of the $eV$ scale but the 
physical mechanism is of the same nature, namely the photon zero point energy becomes variable in space due to spatial variation of photon density of states. Usually in the Casimir effect the force 
is calculated but the method, based on energy calculation, is equivalent. 
\begin{figure}
\includegraphics[width=4cm]{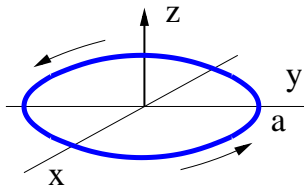}
\caption{\label{fig1}Anomalous state in the form of the circle of the radius $a\sim 10^{-8}cm$. The thickness of the ring thread is on the order of $10^{-11}cm$. The electron momentum along the 
ring produces the angular momentum in the $z$ direction.}
\end{figure}
\subsection{Comments}
\label{anomE}
It happens that the formally singular solution of wave equation does not terminate its story. On short distances $10^{-16}cm$ the natural singularity source enters the game. This source relates to
the generation of electron mass. The fluctuating electromagnetic field washes out the singularity along the $z$ axis turning it into the thread of the subatomically small but finite radius 
$10^{-11}cm$. Within the thread, due to the local reduction of the vacuum energy, the well of $MeV$ scale depth is formed. 

The phenomenon occurs when the $z$ axis coincides with the equilibrium position of the electron at some macroscopic potential extended along that axis. The thread state of the free electron is 
impossible. 

An electron motion in vacuum in the static homogeneous magnetic field $H$ also corresponds to a finite $r_T$. In this case one should substituted $\Omega$ in (\ref{103}) by the cyclotron frequency
$|e|H/mc$. According to (\ref{103}),
\begin{equation}
\label{107a}
r_T\simeq 0.26\sqrt{\ln\frac{4.39\times 10^{9}}{H(T)}}\times 10^{-11}(cm).
\end{equation}
At $H=1T$ the radius $r_T\simeq 1.23\times 10^{-11}cm$. Therefore anomalous electron states in magnetic field in vacuum are possible as in condensed matter. The spectrum of these state is 
continuous (no transverse quantization) and they can be bound with the binding energy of the $MeV$ order. So the electron anomalous states in a magnetic field substantially differ from Landau ones 
\cite{LANDAU5}. 

Usually subatomic physics deals with nuclear and particle phenomena of scales well below the Bohr radius. It appears that electron states of a subatomic size are possible. They are localized at 
positions separated from nuclei. Due to short distances these states relate to $MeV$ energies. So the origin of electron $MeV$ energies in condensed matter is paradoxical solely at first sight. 
\begin{figure}
\includegraphics[width=4cm]{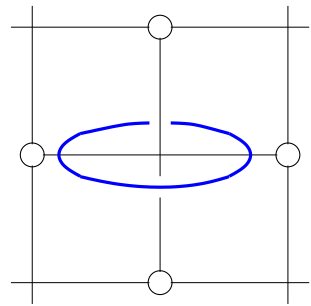}
\caption{\label{fig2} Ring position inside the lattice of a solid. The typical distance between lattice sites is $a_0$.}
\end{figure}
\section{THREAD SHAPE}
\label{shape}
The thread can exist solely along a valley of equilibrium electron positions in some potential. These positions are generally along a curve in three-dimensional space. A small deviation $r$ of the 
thread from that curve costs energy $m\Omega^2r^2/2$ where $\vec r$ is the direction locally perpendicular to the curve. 

In metals a potential along the above valley can be created by a redistribution of conduction electrons related to the energy pay of $\sim 1eV$. But the energy gain, due to electrons in the $MeV$ 
well, is of the order of $1MeV$. Therefore existence of thread state in solids is real. The thread may be of various shapes and lengths. For example, the thread can be restricted by two lattice 
sites taking the position between them along a minimum of the electrostatic potential created by lattice sites and redistributed electrons. 

Electrons in a solid can be redistributed in various manners providing various curves for the equilibrium valley. Suppose this curve to locally deviate from the straight line along the $z$ axis and 
it becomes at $\vec r=\vec u(z)$ where $r_T\ll |\vec u|$. If to take new variables $\{\vec r-\vec u,z\}$ in equation (\ref{M5}), then in the new variables 
\begin{equation}
\label{M15}
-\nabla^2F\rightarrow -\nabla^2F-\left(\frac{\partial u_x}{\partial z}\right)^2\frac{\partial^2F}{\partial x^2},
\end{equation}
where the vector $\vec u$ has the $x$ component only and its $z$ dependence is weak. The evaluation of the second term, as in Sec.~\ref{anomD}, shows that it corresponds to the enhancement of the
electron kinetic energy. Analogously an evaluation of the electromagnetic part also results in energy enhancement 
\begin{equation}
\label{M16}
(\nabla\times\vec A )^2\rightarrow (\nabla\times\vec A )^2+\left(\frac{\partial u_x}{\partial z}\right)^2\left(\frac{\partial A_y}{\partial x}\right)^2.
\end{equation}
We see that a deformation of the valley curve costs energy and the preferable shape of the thread state is linear. This recalls a deformation of an elastic string.  

The thread can be in the form of ring shown in Fig.~\ref{fig1}. For this thread, of the thickness $r_T$, in the form of the circle of the radius $a$ the deformation parameter can be estimated as 
$(\partial\vec u/\partial z)^2\sim r_T/a$. This parameter is constant at each point of the circle whose deformation energy becomes $mc^2(\partial\vec u/\partial z)^2\sim mc^2r_T/a$. Local 
deformations of the circle costs a large energy. Therefore the circle is stiff and it is hardly influenced by lattice sites. For example, the $50\%$ compression of the circle along one axis
costs approximately $1MeV$.
\begin{figure}
\includegraphics[width=5.5cm]{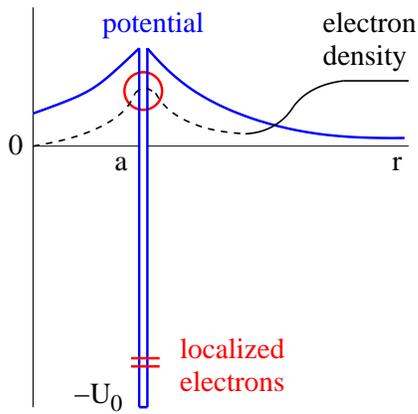}
\caption{\label{fig3} $r$ is distance from the center of the ring of the radius $a$ shown in Fig.~\ref{fig1}. The position $z=0$ is taken. The electron(s), localized deep in the anomalous well of 
the size $(r-a)\sim r_T$, produces the Coulomb barrier for the conduction electron of the energy $E_F$. The circle is of the radius $r_c$. The electron density decays from the thread till 
$(r-a)\sim a_B$. Then it increases going over into the state of conduction electron.}
\end{figure}

To minimize the deformation energy $[a(\AA)]^{-1}keV$ of the thread bent into the circle, its radius $a$ should take its maximal value which is restricted by the distance between lattice sites. 
Such a position is shown in Fig.~\ref{fig2}. Effects of the ring on lattice sites are not important for our purposes. One can consider angular momenta of the ring created by a current along the 
thread. 

Creation of anomalous threads in a solid is described in Ref.~\cite{IVLEV2}. They can be produced either by an irradiation of $keV$ ions of the metallic surface or by an occasional exposure to
radiation. 
\section{ELECTRON STATES OF THE ENERGY $E_F$}
\label{states}
The subatomic potential well of the depth $U_0\sim 1MeV$ is sketched in Fig.~\ref{fig3}. The well acquires electrons from the solid which occupy deeply localized states in the well. This process
is energetically favorable. Transitions of conduction electrons to the well region is restricted by the Coulomb barrier around the thread created by electrons already localized in the well. When 
the number $N$ of electrons in the thread exceeds the certain critical value the barrier prevents further penetration due to a small tunneling probability of electrons of Fermi energy $E_F$. 
\subsection{Origin of electrons localized in the well}
\label{statesA}
Suppose $r=\sqrt{x^2+y^2}$ and $z$ are distances from the center of the ring as in Fig.~\ref{fig1}. The electrostatic potential of the homogeneously charged ring with the total charge $eN$ 
in Fig.~\ref{fig1} has the form at $z=0$ 
\cite{LANDAU4}
\begin{equation}
\label{11}
e\varphi=\frac{2Ne^2}{\pi(r+a)}K\left(\frac{2\sqrt{ar}}{r+a}\right),
\end{equation}
where $K$ is the elliptic integral. At large $r$, $e\varphi\simeq e^2/r$. The electric field is $\vec{\cal E}=-\nabla\varphi$. Close to the thread, $(r-a)\ll a$, at $z=0$
\begin{equation}
\label{12}
e{\cal E}_r\simeq\frac{Ne^2}{\pi a(r-a)}+\frac{Ne^2}{2\pi a^2}\ln\frac{a}{|r-a|}\,.
\end{equation}

The Coulomb barrier (\ref{11}) is sketched in Fig.~\ref{fig3}. Conduction electrons of the Fermi energy leak through the barrier (\ref{11}) via tunneling increasing the number $N$ of electrons 
localized inside the thread. $N$ saturates when the tunneling probability becomes extremely small. To get a general impression about this probability one can approximate $e\varphi$ by $e^2/(r-a)$ 
and then in the approximation of Wentzel, Kramers, and Brillouin \cite{LANDAU5} the tunneling probability is 
\begin{equation}
\label{13}
\frac{1}{t_0}\exp\left(-N\frac{\pi e^2}{\hbar}\sqrt{\frac{2m}{E_F}}\,\right)=\frac{1}{t_0}\exp\left[-\frac{23.2N}{\sqrt{E_F(eV)}}\right],
\end{equation}
where $m$ is the electron mass and $t_0\sim 10^{-15}s$ is a typical atomic time. We consider $s$-wave only. The parameter $(e^2/\hbar)\sqrt{m/E_F}$ in (\ref{13}) is on the order of unity (not a 
semiclassical regime). The coefficient 23.2 is of the numerical origin. If to take a typical $E_F$, the expectation time of filling the well, containing $N$ electrons, is estimated as 
$10^{(8N-15)}s$. For $N=1,2$ the expectation time is not large but for $N=3$ it is years. Therefore the number of electrons, localized in the well, in Fig.~\ref{fig3} is no more than two. 
\subsection{Electron state}
\label{statesB}
The potential energy in Fig.~\ref{fig3} is electrostatic one (\ref{11}) supplemented by the deep well at $r=a$. At this region, inside the circle in Fig.~\ref{fig3}, the electron-photon 
interaction is essential (Sec.~\ref{anom}). Besides deeply localized electrons in the well, there are also states close to the Fermi energy $E_F$. Such state starts with the part, localized close
to the well region ($(r-a)\sim a_B$), and shown by the dashed curve in Fig.~\ref{fig3}. At larger distances that state goes over into the conduction electron shown by the solid curve. 

The state in Fig.~\ref{fig3} is stationary since the lifetime of states in the subatomic well is infinite \cite{IVLEV2,IVLEV3}. This happens since the electron-photon state inside the 
thread is of polaronic type but not of dissipative one when the reservoir is a perturbation. The particular example of such state is studied in \cite{IVLEV3}. One can qualitatively explain why 
photons are not emitted in that state. The electron is connected to the the thread region and is dragged by it. Under photon emission the thread would oscillate increasing the electron kinetic 
energy. This prevents the electron to lose its total energy resulting in non-decaying states. So the $E_F$ electron does not go down in energy at the deep well by quanta emission. 

The ring in Fig.~\ref{fig1} has the angular momentum $\hbar l_z$ due to the circulating current. The underbarrier wave function and its extension from under the barrier in Fig.~\ref{fig3} is 
topological as $\psi\sim\exp(i\varphi l_z)$ where $\varphi$ corresponds to rotation around the $z$ axis. 

At small distances from the ring $(r-a)\lesssim r_c$, within the circle in Fig.~\ref{fig3}, relativistic effects are strong \cite{IVLEV2}. This reminds strong relativistic effects in some atoms 
where instead of $ls$ coupling there is $jj$ one. In our case this is $j_zj_z$ coupling due to cylindrical (circle) symmetry. The crystal field hardly violate the narrow region around the thread
where $j_zj_z$ coupling is formed. This means that for the thin thread circle the energy is characterized by $j_z$ quantum number but not by $s_z$ and $l_z$ separately. The corresponding electron 
state continues from the thread to outside. 

Those quantum mechanical effects, that is without photons participation, do not influence superconducting state or pairing effects. Analogously spin-orbit phenomena do not affect a 
superconducting state as known.
\section{SPIN IMBALANCE STATE}
\label{spin}
In this section we study how the electron state, close to the thread circle in Fig.~\ref{fig3}, goes over into a conduction electron at large distances. We start with the effect of the interaction 
with photons at $(r-a)\lesssim a_B$. 
\subsection{Lamb shift}
\label{spinA}
First of all, we emphasize that the Lamb shift, considered in this section, does not relate to atomic levels. In our case this is an energy shift of electron levels close to $E_F$ caused by the 
Coulomb field of an electron localized deep in the well (in presence of the electron-photon interaction) in Fig.~\ref{fig3}. The associated electron density is plotted in Fig.~\ref{fig3}. 

The description (\ref{101}) and (\ref{102}) of the Lamb shift is referred to $l=0$. When $l\neq 0$ it is better to use the first non-zero term of the perturbation theory for the energy Lamb 
shift \cite{LANDAU3}
\begin{equation}
\label{16}
E_L=-\frac{e^3}{2\pi m^2c^3}\langle\,|\vec s\,(\vec{\cal E}\times\vec p)|\,\rangle, 
\end{equation}
Accounting for the relation $\hbar\vec l=\vec R\times\vec p$ one can obtain from (\ref{16}) 
\begin{equation}
\label{17}
E_L=\frac{e^2\hbar (s_zl_z)}{2\pi m^2c^3}\Big\langle\,\Big|\frac{\vec r}{r^2}\frac{\partial e\varphi}{\partial\vec R}\Big|\,\Big\rangle,
\end{equation}
where only $s_zl_z$ part survives after the spatial average. 

The main contribution to the matrix element in (\ref{17}) comes from the underbarrier wave function in Fig.~\ref{fig3}. Estimating from the second term in (\ref{12}) 
$\partial e\varphi/\partial R\sim e^2/a^{2}_{B}$, we obtain
\begin{equation}
\label{18}
E_L=-2(s_zl_z)\varepsilon_L\,,\hspace{0.3cm}\varepsilon_L\sim\frac{me^4}{\hbar^2}\left(\frac{e^2}{\hbar c}\right)^3.
\end{equation}

The usual spin-orbit term $\vec l\vec s$ is a part of the Hamiltonian related to the wave equation \cite{LANDAU3}. That term is time-reversal and therefore it does not influence superconducting 
state. 

The Lamb shift result (\ref{17}) also looks as one originated from some correction to the potential energy as in the case of spin-orbit. But the Lamb shift phenomenon is not reduced to a 
correction of the potential energy. The point is that in formation of the result (\ref{17}) virtual photons are involved \cite{LANDAU3}. Due to this the motion is not characterized by an
electron momentum only which changes sign under the time reverse. Therefore opposite spins, referred to the split (\ref{18}), produce the depairing effect on superconductivity. 
\begin{figure}
\includegraphics[width=7cm]{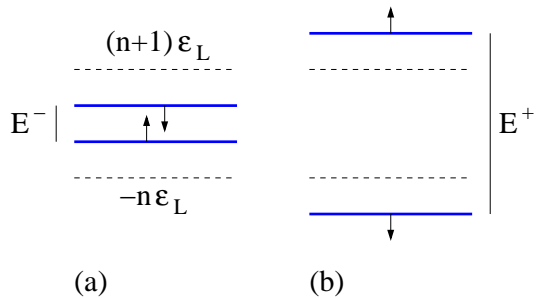}
\caption{\label{fig4}Scheme of the spin imbalance state close to the ring. With the magnetic field energy differences (a) $E^-$ and (b) $E^+$ between electrons of opposite spins are shown by 
arrows. Without the magnetic field the energy difference is the same, $(2n+1)\varepsilon_L$, for the both cases (broken lines).}
\end{figure}
\subsection{Spin imbalance states}
\label{spinB}
The total angular momentum $j_z=n+1/2$ ($n\geq 0$) can be realized in two ways 
\begin{equation}
\label{20}
j_z=n+\frac{1}{2}\Rightarrow
\begin{cases}
\{\downarrow,\,l_z=n+1\}\hspace{0.4cm} E_L=(n+1)\varepsilon_L\\
\{\uparrow,\,l_z=n\}\hspace{1.0cm} E_L=-n\varepsilon_L.
\end{cases}
\end{equation}
Arrows up and down show spin directions along the $z$ axis. The values $2l_zs_z=-n-1$ and $2l_zs_z=n$ produce Lamb 
energies in (\ref{18}). Analogously, the total angular momentum $j_z=-n-1/2$ ($n\geq 0$) can be realized also in two ways
\begin{equation}
\label{21}
j_z=-n-\frac{1}{2}\Rightarrow
\begin{cases}
\{\uparrow,\,l_z=-n-1\}\hspace{0.4cm} E_L=(n+1)\varepsilon_L\\
\{\downarrow,\,l_z=-n\}\hspace{1.0cm} E_L=-n\varepsilon_L.
\end{cases}
\end{equation}
The energy split (broken lines in Fig.~\ref{fig4}) between pair of states in (\ref{20}) or (\ref{21}) can be written in the form
\begin{equation}
\label{22}
\Delta E_L=(2n+1)\varepsilon_L
\end{equation}
at any integer $n$.

We see that under spin-orbit interaction the level with the fixed $j_z$ was double degenerated with $l_z=j_z\pm 1/2$. The electron-photon interaction removes this degeneracy. That is similar 
to hydrogen atom where spin-orbit interaction remains degenerated two states with the same $j$ but different $l=j\pm 1/2$. The electron-photon interaction removes the degeneracy in hydrogen atom 
(Lamb shift) \cite{LANDAU3}. 

As shown in Appendix A, the wave function in Fig.~\ref{fig3} is reasonably localized close to the ring and $n$-dependence of $\varepsilon_L$ is weak.
\begin{figure}
\includegraphics[width=5cm]{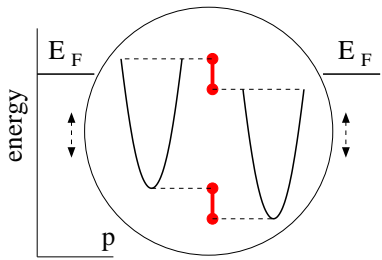}
\caption{\label{fig5} Scheme of the spin-imbalance state in the volume of a metal. The states of opposite spins are separated in energy by ``stiff dumbbells''. For convenience these subsystems
are drawn shifted in momentum $p$. The entire system (within the circle) oscillates, along the energy axis, under interaction with phonons keeping the same energy split (\ref{22}) between opposite 
spin subsystems. These oscillations are denoted by dashed arrows.}
\end{figure}
\subsection{Why small thread rings strongly influence conduction electrons}
\label{spinC}
The above classification is applicable to the region of the Bohr radius size near the thread ring. At $(r-a)\lesssim a_B$ electron states with opposite spins are split in energy according to
(\ref{22}). After coming out from under the barrier in Fig.~\ref{fig3} electrons are scattered by lattice sites and impurities. These processes are elastic and therefore the electron keeps the
energy split (\ref{22}) for opposite spins. In the volume electrons are no more described by orbital quantum numbers but instead by momenta in the lattice $\vec p$. Electrons of opposite spin 
relate to the energies $\varepsilon(\vec p)+(n+1)\varepsilon_L$ and $\varepsilon(\vec p)-n\varepsilon_L$, where $\varepsilon(\vec p)$ is the energy spectrum in the lattice.  

This is shown in Fig.~\ref{fig5} where ``stiff dumbbells'' separate in energy electrons of opposite spins. The energy split (\ref{22}) for opposite spin directions remains stiff in the crystal 
lattice within the spin-orbit relaxation length $a_0(\hbar c/e^2)^2$ which is approximately a few microns. Within this scale there is no equilibrium between Fermi levels of subsystems with opposite 
spins as shown in Fig.~\ref{fig5}. The number of spin up and spin down electrons are the same. When the mean distance between thread rings is shorter than the spin-orbit length, such spin imbalance 
state exists in the entire volume. Thread rings, distributed in the volume, are the driving force for spin imbalance state. 

Inelastic processes in a metal, resulted from electron-phonon effects, are characterized by the uncertainty $T^3/(\hbar\omega_D)^2$ of the electron energy (imaginary part of the spectrum). Here
$\omega_D$ is the Debye frequency. Those processes can be interpreted as oscillations, along the energy axis, of the entire system (the circle in Fig.~\ref{fig5}). This is shown by dashed arrows. 
Under these oscillations the energy split between opposite spin subsystems in Fig.~\ref{fig5} remains the same. The total spin imbalance state is a superposition of ones characterized by energy 
splits (\ref{22}) with various $n$.

There is a difference in states in the bulk generated by thread circles and ones in the usual scattering by impurities. The latter hardly influence electron states in the volume. Atomic size rings 
also can be treated as impurities. But the essential feature of such impurities is the inner structure of them with the subatomic region within the thread. The state parameters (spin imbalance), 
formed on that small scale, are stiff and transformed through a relatively transparent barrier to the bulk. 

There is an analogy with the usual impurity scattering when the impurity also has an inner structure: a discrete energy level. In this case the scattering amplitude of particles, with the energy
close to resonance one (Wigner resonance scattering), is anomalously large \cite{LANDAU5}. 
\subsection{Influence of the magnetic field}
\label{spinD}
The action of the external magnetic field on the spin imbalance state is not described by the Zeeman term $\mu_B(\vec l+2\vec s)\vec H$ (not by a $g$-factor), as for an atom, since in the volume 
there is no orbital quantum number $l$. Here $\mu_B=|e|\hbar/2mc$ is the Bohr magneton. The orbital part goes over into the diamagnetic one, $(e/mc)\vec p\vec A$, in the volume. Due to impurity 
scattering the diamagnetic part provides the continuous spin-independent contribution to the total spectrum. This is not significant for our purposes. Therefore the influence of the magnetic field 
can be accounted for through the paramagnetic part $2\mu_B\vec s\vec H$ only.

Suppose the applied magnetic field $H$ to be directed along the $z$. The paramagnetic energy $2\mu_Bs_zH$ enters the game. To be specific suppose $H>0$. Then for the cases (\ref{20}) 
(Fig.~\ref{fig4}(a)) and (\ref{21}) (Fig.~\ref{fig4}(b)) the level splits are
\begin{equation}
\label{23}
E^{\mp}_n=(2n+1)\varepsilon_L\mp 2\mu_BH,
\end{equation}
where two energies refer to the states with $s_z=\pm 1/2$. 

When in layered compounds thread circles are in $ab$ planes, the magnetic field in Eq.~(\ref{23}) is one directed along the $c$ axis. When the circle plane is perpendicular to $ab$ plane, $H$ in
Eq.~(\ref{23}) corresponds to one in the $ab$ plane. 
\section{EFFECT ON COOPER PAIRING}
\label{cooper}
The Cooper pair can be formed by the electron with the energies $\varepsilon (-\vec p\,)-E_F-n\varepsilon_L+\mu_BH$ and $\varepsilon (\vec p\,)-E_F+(n+1)\varepsilon_L-\mu_BH$ (Fig.~\ref{fig4}(a)). 
The former refers to the state (denoted as $\uparrow$) with the spin superposition along the $z$ axis and along $\vec H$. The latter $(\downarrow)$ relates to mutually inverted spins. As plotted
in Fig.~\ref{fig5}, the Fermi levels of subsystems with opposite spins are also shifted by the same energy. 

Pairing of those spin imbalance states correspond to the order parameter $\Delta^{\downarrow\uparrow}$. Analogously the component $\Delta^{\uparrow\downarrow}$ is formed, according to 
Fig.~\ref{fig4}(b). Above $T_c$ instead of order parameter there are fluctuation propagators satisfying equations (see Appendix B)
\begin{eqnarray}
\label{24}
\left[\frac{i\pi}{8T}(-\hbar\omega +E^{-}_{n})+\frac{T-T_c}{T}+\xi^2 k^2\right]\Delta^{\downarrow\uparrow}_n=0\\
\label{25}
\left[\frac{i\pi}{8T}(-\hbar\omega +E^{+}_{n})+\frac{T-T_c}{T}+\xi^2 k^2\right]\Delta^{\uparrow\downarrow}_n=0,
\end{eqnarray}
where $\xi\sim\hbar v_F/T_c$ is the coherence length. These propagators differ from usual ones \cite{VARL} by non-zero energies $E^{\mp}_{n}$. Due to gauge invariance it is impossible to eliminate 
$E^{\mp}_{n}$ in those equations by choosing proper phases of $\Delta_n$. This is due to the difference in Fermi levels of two subsystems with opposite spins in Fig.~\ref{fig5} where paramagnetic 
shifts are included.  

At first sight, one can choose the new gauge $\Delta\rightarrow\exp(i\chi)\Delta$ to compensate $E^{\pm}_{n}$, or a part of them, by $i\hbar\dot\chi$. But in this case additional non-stationary 
terms appear in the formalism of the diagram technique near $T_c$ and the final result for resistance remains the same as for $\chi=0$. 
\begin{figure}
\includegraphics[width=7cm]{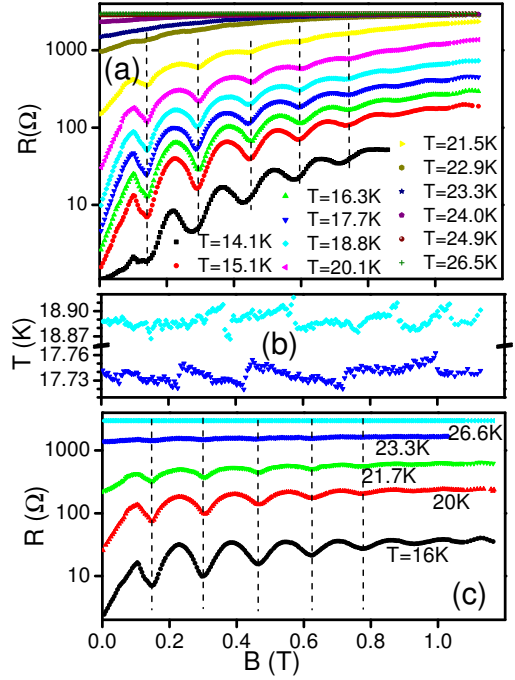}
\caption{\label{fig6} Magnetoresistance curves in ${\rm Sr}_{0.88}{\rm La}_{0.12}{\rm CuO}_2$ sample \cite{MIL}. The universal positions of maxima 
correspond to Eq.~(\ref{26}) including ``1/2\,''. Each peak can be marked by $n=0,1,2..$. (a) The case of $B^{\bot}$. (b) Temperature control. 
(c) $B^{\|}$ curves in same sample.}
\end{figure}

The specificity of spin imbalance state, resulting in the propagators (\ref{24}) and (\ref{25}), is shifted Fermi energies of subsystems with opposite spins. This state is supported by rings
distributed in the volume. In the usual equilibrium metal Fermi levels of subsystems with opposite spins coincide (the length of the upper ``dumbbell'' in Fig.~\ref{fig5} is zero). In this case 
the Zeeman terms in Eqs.~(\ref{24}) and (\ref{25}) are absent. Instead there is the depairing term $(\mu_BH/T)^2$. See also \cite{MAKI,ICH}. Formally this follows from Appendix B, where in 
Eq.~(\ref{B2}) the arguments of tangents are not shifted by $E_{1,2}$ (coinciding Fermi levels). 
\subsection{$R(H)$ oscillations}
\label{cooperA}
Above $T_c$ the electric resistance differs from its value in the normal metal by the fluctuation part which is determined by fluctuation 
propagators (\ref{24}) and (\ref{25}) \cite{VARL}. The measured $R(H)$ is a sum on spin directions and depends on all propagators 
$\Delta^{\downarrow\uparrow}_n$ and $\Delta^{\uparrow\downarrow}_n$. Contributions of propagators to resistance are negative. With finite
energy shifts $E^{\mp}_{n}$ these contributions are enhanced by the factor proportional to $|E^{\mp}_{n}|/\Delta T$ (compare with \cite{VARL}). 
This factor essentially increases the fluctuation contribution since $\Delta T$ is a small width related to the fluctuation region near $T_c$. At
$E^{\mp}_{n}=0$ the factor equals unity which is the conventional case of non-shifted energy \cite{VARL}. Well below $T_c$ any oscillation effect, related to the condition $E^{\pm}_{n}=0$, is 
small since it is determined by $E^{\pm}_{n}/T_c$.

Therefore the most weak contribution of $|E^{-}_{n}|$ occurs when that value is zero. This condition (mostly restored normal resistance) corresponds to pronounced maxima on the $R(H)$ curve. 
Positions $H_n$ of maxima of $R(H)$, corresponding to the condition $|E^{-}_{n}|=0$, are
\begin{equation}
\label{26}
H_n=\left(\frac{1}{2}+n\right)\Delta H,\hspace{0.8cm}\Delta H=\frac{\varepsilon_L}{\mu_B} 
\end{equation}
with all integer $n$. With the choice $H<0$ the energy $E^{+}_{n}$ is involved instead of $E^{-}_{n}$ and the condition (\ref{26}) remains the same for integer $n$ of any sign. 

The calculated period of $R(H)$ oscillations is $\Delta H\simeq 0.18~T$. We use the approximate estimate (\ref{18}). One should emphasize that the oscillations of $R(H)$ in the fluctuation region 
are due to the periodic in $H$ coincidence of Fermi energies for opposite spins.

Experimental magnetoresistance curves for different orientations of the magnetic field are shown in Fig.~\ref{fig6}. Positions of maxima of 
resistance coincide very good with the condition (\ref{26}). First, the periodicity follows from the theory. Second, the observed periodicity 
$0.155T$ is close to calculated one. Third, even ``1/2'' in Eq.~(\ref{26}) corresponds to observations.

We see that the positions of maxima on the oscillatory curve $R(H)$ are determined by paramagnetic effects related to the condition $E^{\mp}_{n}=0$. To analyze the entire shape of the curve (for 
example the steady slope) one should include also diamagnetic effects. 

The periodicity of peak positions in Fig.~\ref{fig6} is within the $5\%$ uncertainty. On the other hand, the matrix element (\ref{17}) depends on the wave function outside the deep well in 
Fig.~\ref{fig3}. In turn, that wave function depends on $l_z$ since usually at larger orbital momentum the wave function is localized at larger distances from the center. This would reduce the 
matrix element (\ref{17}) at larger $l_z$. Therefore $\varepsilon_L$ in Eq.~(\ref{18}), strictly speaking, depends on $n$ violating the periodicity on magnetic field. 

But in our case the electron distribution cannot be shifted toward larger distances under the increase of $l_z$ due to the fixed position of the well at $r=a$. As shown in Appendix A, the electron
distribution is localized close to $r=a$ which results in a weak dependence of $\varepsilon_L$ (and therefore of $\Delta H$) on $n$.

In layered compounds thread circle planes are oriented in two different ways: in $ab$ planes and perpendicular to them. The former rings are responsible for the periodic $R(H)$ when $\vec H$ is
perpendicular to $ab$ planes. The latter rings determine $R(H)$ when $\vec H$ in in $ab$ planes (Sec.~\ref{spinD}). These two possibilities are presented in Fig.~\ref{fig6}. 
\section{DISCUSSION}
\label{disc}
We study the phenomenon which does not fall into the set of known mechanisms. Universality of $R(H)$ periodicity  with respect to magnetic field orientation and a very weak dependence on material 
turned us to look for a different scenario. Likely a subatomic mechanism, which is material independent, could relate to the phenomena observed. The non-trivial issue in the whole story is the 
introduction of electron-photon subatomic mechanism.

It is unusual that subatomic phenomenon plays a substantial role into condensed matter physics. We emphasize that the subatomic mechanism involved is not referred to nuclear and particle phenomena 
but to electron ones. In this paper the subatomic electron mechanism is proposed which explains the unconventional experimental results. In that mechanism the spatial scale of the electron system 
is of the Compton length $\hbar/mc\sim 10^{-11}cm$. This is $10^{3}$ times less than the atomic size. 

The basis for that is a state where the electron density is formally singular on the certain line. This is possible according to quantum mechanics of electron and arguments stemming from mechanisms 
of its mass generation. Due to the interaction with photons the electron ``vibrates'' leading to smearing of that singularity within the thread of the subatomically small radius $10^{-11}cm$. This 
anomalous electron state is accompanied by a well of the depth $\sim 1MeV$ localized within the narrow thread region. This energy scale is unexpected in condensed matter. The origin of the well is 
due to a local reduction of electromagnetic zero point energy. 

The thread is not necessary linear. The subatomically thin thread can be in the form of a ring of the interatomic radius. In a metal the role of such rings is unusual due to orbital momenta of the 
ring along the $z$ axis (perpendicular to the ring plane). The substantial issue is the subatomic smallness of the thread thickness. Due to this, inner properties on such scale do not depend on the
crystal field.  

Close to the thread (on the Compton length) relativistic effects are strong resulting in $j_zj_z$ coupling analogous to $jj$ coupling in some atoms. So relativistic quantum mechanical states are 
marked by $j_z$. At atomic distances from the thread, due to the interaction with photons, the state of the Fermi energy is split by two ones, with $l_z=j_z+1/2$ and  $l_z=j_z-1/2$. This is similar 
to the Lamb split in hydrogen atom where, instead of $z$ component, total momenta are involved due to spherical symmetry.  

The electron of the Fermi energy probes the well of the $MeV$ depth close to the thread. Those states, with energy split for opposite spins, are continued across the barrier to larger distances 
going over into conduction electrons close to $E_F$ in energy. For this reason, that narrow stiff region plays a role of a boundary condition for conduction electrons moving in $eV$ crystal field. 
This keeps electrons with opposite spins in the volume to be separated by discrete Lamb energies. 

Such spin imbalance state in the volume is relaxed, due to spin-orbit effects, on the distance of a few microns. But when the mean distance among rings is smaller, the spin imbalance state exists 
in the entire volume of the metal. Note that usual impurity atoms result in simple scattering of conduction electrons with continuous energies. 

That spin imbalance state of conduction electrons, with discrete energy splits for opposite spins, influences the Cooper pairing condition. The resistance in Ref.~\cite{MIL} was measured close 
to $T_c$ in the fluctuation region. Under this condition the fluctuation correction to the resistance of normal metal is determined by the fluctuation propagators. They, in turn, depend on energy 
shifts of different spin states. These energy shifts can be subsequently turned to zero by the external magnetic field. Therefore the resulting $R(H)$ dependence becomes oscillating as in 
experiments. There is a good coincidence of the experimental, Fig.~\ref{fig6}, and theoretical, Eq.~(\ref{26}) (including ``1/2''), positions of magnetoresistance peaks. 

In a solid threads can be created during sample preparation or through exposure to radiation. For example, ions of $keV$ energy, bombarding the sample, have the wave length $\sim 10^{-11}cm$ and 
produce charge density of the same scale after reflections from lattice sites. The matrix element of that perturbation between a conduction state and anomalous one is not small. Samples with 
identical materials and geometry, fabricated under different conditions (at different labs), can exhibit oscillatory magnetoresistance or not regarding threads generation in a sample in the process 
of fabrication. 

The observed oscillatory magnetoresistance is an implicit manifestation of anomalous states. One can compare this  with observation of X-ray laser pulses from the ``dead'' sample during 20 hours 
(see \cite{IVLEV2}). In that case there is an explicit manifestation of anomalous states. 

Anomalous electron states in vacuum in a magnetic field are possible as in condensed matter. The spectrum of these state is continuous (no transverse quantization) and they can be bound with the 
binding energy of the $MeV$ order. So the electron anomalous states in a magnetic field substantially differ from Landau ones.
\section{CONCLUSIONS}
The observed universal oscillations of magnetoresistance are associated with subatomic states inside the superconductor. Such states are the subatomically thin ($10^{-11}cm$) threads in the form of 
the rings of the interatomic radius ($10^{-8}cm$). In the thread region the subatomic potential well of the $MeV$ depth is formed which is unusual in condensed matter physics. From thread regions 
electron states continue to the volume producing there spin imbalance state. This state is probed in the measurements. Calculated universal positions of peaks $(n+1/2)\Delta H$ ($n=0,1,2...$) on 
the $R(H)$ curve are in a good agreement with measurements. 

\acknowledgments
I thank M. Kunchur, J. Knight, and R. Prozorov for discussions. This work was supported by CONACYT through grant number~237439.
\appendix
\section{DEPENDENCE OF $\varepsilon_L$ ON $n$}
Below we evaluate the form of the underbarrier wave function which is responsible for the parameter $\varepsilon_L$ (\ref{18}). This function describes the electron outside the circle in 
Fig.~\ref{fig3}. Inside that circle the electron-photon hybridization occurs and a description by the wave equation is not valid. That circle (thread) region plays a role of a boundary condition 
for the outside region. For simplicity one can consider the constant potential energy instead of the Coulomb one (\ref{11}). In this case it is convenient to use the dimensionless Schr\"{o}dinger
equation 
\begin{equation}
\label{A1} 
-\nabla^2\psi+\psi=c\delta(z)\delta(r-a)\exp(il_z\varphi),
\end{equation}
where the right-hand side is analogous to $\delta(\vec r)$ for the linear thread along the $z$ axis in Sec.~\ref{anomC}. The coordinates relate to Fig.~\ref{fig1}. The constant $c$ stays for the 
normalization of the wave function. The dimensionless $a\sim 1$ corresponds to the Bohr radius $a_B$. In Eq.~(\ref{A1}) the components $\vec r=\{r\cos\varphi,\,r\sin\varphi\}$ are used. 

It is easy to show that the Fourier component of the wave function is
\begin{equation}
\label{A2} 
\psi_k=2c\pi a\exp\left(-\frac{i\pi l_z}{2}+il_z\varphi_1\right)\frac{J_{l_z}(ka)}{k^{2}_{z}+k^2+1}\,,
\end{equation}
where 
\begin{equation}
\label{A3} 
J_{n}(v)=\int^{2\pi}_{0}\frac{d\theta}{2\pi}\exp(-in\theta+iv\sin\theta)
\end{equation}
is the Bessel function and $\vec k=\{k\cos\varphi_1,\,k\sin\varphi_1\}$. As follows from (\ref{A2}), the wave function is
\begin{eqnarray}
\label{A4} 
&&\psi(\vec r,z)=ca\exp(il_z\varphi)\\
\nonumber
&&\times\int^{\infty}_{0}\frac{kdk}{2\sqrt{1+k^2}}\exp\left(-|z|\sqrt{1+k^2}\right)J_{l_z}(ka)J_{l_z}(kr).
\end{eqnarray}\\
\subsection{Close to the axis of the ring}
At $r\ll a$ one can use the asymptotics $J_n(v)\simeq (v/2)^n/n!$ at small arguments for $J_{l_z}(kr)$. The result is
\begin{eqnarray}
\label{A5} 
&&\psi(\vec r,z)=\frac{c}{2l_z!}\,\exp\left(i\varphi l_z\right)\,\left(\frac{r}{2a}\right)^{l_z}\\
\nonumber
&&\times\int^{\infty}_{0}\frac{v^{l_z+1}dv}{\sqrt{v^2+a^2}}\,J_{l_z}(v)\exp\left(-\frac{|z|}{a}\sqrt{v^2+a^2}\right).
\end{eqnarray}
\subsection{Close to the thread}
In the limit $(r-a),\,z\ll a$ large $k$ in (\ref{A4}) are essential. With the asymptotics
\begin{equation}
\label{A6} 
J_{n}(v)\simeq\sqrt{\frac{2}{\pi v}}\cos\left(v-\frac{\pi n}{2}-\frac{\pi}{4}\right),\hspace{0.3cm}1\ll v
\end{equation}
it follows from (\ref{A4})
\begin{equation}
\label{A7} 
\psi(\vec r,z)=\frac{c}{4\pi}\exp(i\varphi l_z)\ln\frac{1}{(r-a)^2+z^2}\,.
\end{equation}
The wave function logarithmically diverges close to the thread as it should be (Sec.~\ref{anomC}).
\subsection{Far from the ring}
In the case $a\ll r$ in Eq.~(\ref{A4}) $kr\sim 1$ and $ka\ll 1$. According to these limits, 
\begin{eqnarray}
\label{A8} 
&&\psi(\vec r,z)=\frac{c}{2l_z!}\,\exp\left(i\varphi l_z\right)\,\frac{\exp(-|z|)}{r^2}\\
\nonumber
&&\times\left(\frac{a}{2r}\right)^{l_z}\int^{\infty}_{0}v^{l_z+1}J_{l_z}(v)dv.
\end{eqnarray}
\subsection{Dependence of $\varepsilon_L$ on $n$}
One can conclude from Eqs.~(\ref{A5}) and (\ref{A8}) that the electron density $|\psi|^2$ strongly decays with the distance $(r-a)$ from the ring. In other words, it is localized close to the ring 
since, after the adjustment of the constant $c$,
\begin{equation}
\label{A9} 
\int |\psi|^2d^2rdz=1
\end{equation} 
for all $l_z$.

The Lamb energy $\varepsilon_L$ (\ref{18}) is determined by the formal matrix element (\ref{17}) where $|\psi|^2$ is integrated with the electric field. This field contains the part (the first
term in (\ref{12})) which is odd with respect to $(r-a)$ and therefore weakly contributes to the integral in (\ref{17}). The second term in (\ref{12}) is even with respect to $(r-a)$ and slightly
varies close to $r=a$ where $|\psi|^2$ is mainly localized. Therefore, due to the condition (\ref{A9}), the matrix element (\ref{17}) hardly depends on $l_z$. For this reason, $\varepsilon_L$
and $\Delta H$ in (\ref{26}) weakly depend on $n$. 
\section{FLUCTUATION PROPAGATOR IN SPIN IMBALANCE STATE}
Suppose that in Fig.~\ref{fig5} the left spectrum refers to spin $(\downarrow)$ and the right one to to spin $(\uparrow)$. Fluctuation propagators (\ref{24}) and (\ref{25}) depend on $\omega$ and
$k$. As the first step, suppose $k=0$. We consider the phonon mechanism of pairing. The final result hardly depends on this choice. Then the propagator is determined by the equation 
\cite{VARL,KOP3}
\begin{equation}
\label{B1} 
\left(\frac{1}{|g|}+\int^{\infty}_{-\infty}\frac{d\varepsilon}{4\pi i}\,Q_{\varepsilon}\right)\Delta^{\downarrow\uparrow}_n=0,
\end{equation} 
where $g$ is the electron-phonon constant and 
\begin{eqnarray}
\label{B2} 
\nonumber
\int d\xi_p\bigg[\left(G^{-R}_{\varepsilon\,p}\right)^{\downarrow\downarrow}\left(G^{+R}_{\varepsilon-\hbar\omega\,p}\right)^{\uparrow\uparrow}\tanh\frac{\varepsilon-\hbar\omega-E_2}{2T}\\
-\tanh\frac{\varepsilon-E_1}{2T} \left(G^{-A}_{\varepsilon\,p}\right)^{\downarrow\downarrow}\left(G^{+A}_{\varepsilon-\hbar\omega\,p}\right)^{\uparrow\uparrow}\bigg]=Q_{\varepsilon}\,.
\label{B2}
\end{eqnarray} 
Here retarded and advanced Green's functions are
\begin{eqnarray}
\label{B3} 
\left(G^{-R,A}_{\varepsilon\,p}\right)^{\downarrow\downarrow}=(\varepsilon-\xi_p-E_1\pm i\delta)^{-1}\\
\label{B3a}
\left(G^{+R,A}_{\varepsilon\,p}\right)^{\uparrow\uparrow}=(\varepsilon+\xi_p+E_2\pm i\delta)^{-1},
\end{eqnarray} 
where $\xi_p=\varepsilon(p)-E_F$ (we suppose the isotropic particle spectrum $\varepsilon(p)$) and the positive $\delta$ is small. In equations (\ref{B3}) and (\ref{B3a})
\begin{equation}
\label{B4} 
E_1=(n+1)\varepsilon_L-\mu_BH,\hspace{0.5cm}E_2=-n\varepsilon_L+\mu_BH.
\end{equation} 
In Eq.~(\ref{B2}) there is also the cross term, containing $G^{-R}_{\varepsilon}G^{+A}_{\varepsilon-\hbar\omega}$, but it does not contribute in our case \cite{KOP3}. 

Performing the pole integration on $\xi_p$ in (\ref{B2}), we obtain
\begin{equation}
\label{B5} 
Q_{\varepsilon}=-\frac{2\pi i}{R_{\varepsilon}}\tanh\frac{\varepsilon-\hbar\omega-E_2}{2T}-\frac{2\pi i}{R^{*}_{\varepsilon}}\tanh\frac{\varepsilon-E_1}{2T},
\end{equation} 
where $R_{\varepsilon}=2\varepsilon-\hbar\omega-E_1+E_2+2i\delta$. 

The integration in (\ref{B1}), with the expression (\ref{B5}), consists of the pole part and the contribution of large $\varepsilon>T$. According to that, the equation (\ref{B1})reads
\begin{equation}
\label{B6} 
\left[\frac{1}{|g|}-\int^{\omega_D}_{0}\frac{d\varepsilon}{\varepsilon}\tanh\frac{\varepsilon}{2T}+\frac{i\pi}{8T}(E_1-E_2-\hbar\omega)\right]\Delta^{\downarrow\uparrow}_n=0
\end{equation} 
with the upper $\omega_D$ cut off. In the second term in (\ref{B6}) we neglected $\hbar\omega$, $E_1$, and $E_2$ which are small compared to $T$. 

The integral in (\ref{B6}) is evaluated as $\ln\omega_D/T$. Due to the relation for the phonon model $1/|g|=\ln\omega_D/T_c$, the first two terms in (\ref{B6}) are $\ln T/T_c$ which is 
$(T-T_c)/T_c$ close to $T_c$. Now it follows from (\ref{B6}) 
\begin{equation}
\label{B7}
\left[\frac{i\pi}{8T}(-\hbar\omega +E^{-}_{n})+\frac{T-T_c}{T}\right]\Delta^{\downarrow\uparrow}_n=0,
\end{equation}
where we use the relation $E_1-E_2=E^{-}_{n}$. 

Before we consider the harmonics of $\Delta^{\downarrow\uparrow}_n$ with $k=0$. It is not difficult to account for finite $k$. After the routine procedure with the substitution in (\ref{B3a}) 
$\xi_p\rightarrow\xi_p-\vec v_F\vec k$, one obtains Eq.~(\ref{24}) with the coherence length $\xi\sim\hbar v_F/T_c$. Analogously one can derive Eq.~(\ref{25}).

\end{document}